\documentclass[aps,prx,twocolumn,amsmath,amssymb,superscriptaddress,floatfix]{revtex4-1}

\usepackage{graphicx}
\usepackage{multirow}
\usepackage{color}
\usepackage{bm}
\usepackage{hyperref}

\renewcommand{\Vec}[1]{{\bm{#1}}}

\def\t#1{\textrm{#1}}
\def\ket#1{|#1\rangle }
\def\bra#1{\langle #1 |}
\def\braket#1{\langle #1 \rangle}
\def\n{\nonumber \\ }

\begin{document}
\title{
Topological aspects of nonlinear excitonic processes in noncentrosymmetric crystals
}
\author{Takahiro Morimoto}
\affiliation{Department of Physics,
University of California, Berkeley, CA 94720}
\author{Naoto Nagaosa}
\affiliation{RIKEN Center for Emergent Matter Science 
(CEMS), Wako, Saitama, 351-0198, Japan}
\affiliation{Department of Applied Physics, The University of 
Tokyo, Tokyo, 113-8656, Japan}

\date{\today}

\begin{abstract}
We study excitonic processes second order in the electric fields in noncentrosymmetric crystals. We derive formulas for shift current and second harmonic generation produced by exciton creation, by using the Floquet formalism combined with the Keldysh Green's function method.  
It is shown that (i) the steady dc shift current flows by exciton creation without dissociation into free carriers and (ii) second harmonic generation is enhanced at the exciton resonance.
The obtained formulas clarify topological aspects of these second order excitonic processes which are described by Berry connections of the relevant valence and conduction bands.
\end{abstract}

\maketitle

\section{Introduction}
Nonlinear optical processes in solids are the important subject in 
condensed matter physics, which are also of crucial importance for 
applications\cite{Bloembergen,Boyd,Toyozawa}. 
In particular, noncentrosymmetric crystals host the second 
order processes in electric fields such as shift 
current, optical rectification and second harmonic generation (SHG). 
We have recently 
revealed that these second order optical processes are topological in nature and are described 
by the Berry connections of Bloch wavefunctions of conduction and 
valence bands for non-interacting electrons~\cite{Morimoto}. 
However, the role of Coulomb interaction is often essential in the electronic 
processes in insulating solids. 
In particular, excitons, bound pairs of electron and 
hole via Coulomb interaction, are relevant to these processes by enhancing these nonlinear effects in many cases~\cite{Toyozawa}
Therefore, it is an important issue to study the interaction effect on the 
topological nonlinear optical effects, which we address in this paper.

One example of the second order optical effects is the photocurrent, which is also relevant to the solar cell action. 
Light irradiation creates the electrons and 
holes in the crystal, which are often bound to form neutral excitons.
It is believed that the excitons cannot contribute to the steady dc current. 
If this is the case, the dissociation of the excitons into free electrons and holes 
is essential to produce the dc photocurrent. This process is usually achieved by the potential 
gradient at p-n junction or by the applied electric field. The efficiency of the solar cell 
action is largely determined by the probability of the thermally activated dissociation 
process which is competing with the annihilation of the excitons \cite{Onsager}. 
However, we show below that the excitons can support the dc current under 
the steady light irradiation 
due to the geometrical nature of the Bloch wavefunctions.  

In the past decades, it has been recognized that the free carriers are not necessarily 
needed for the current. 
The representative example is the polarization current in ferroelectrics  \cite{Resta}.
In an insulator with a band gap, electrons occupying the valence band can support 
current corresponding to the time derivative of the polarization. 
This current is characterized by the Berry phase of the valence electrons. 
Specifically, the Berry phase is related to the ``intra-cell'' 
coordinates, i.e., the band dependent shift of the wavepacket made from the Bloch 
wavefunctions. 
This shift of the electrons described by the Berry phase is the origin of the electric 
polarization that leads to the polarization current. 
However, the polarization current cannot be a steady dc current.
Since the polarization current usually appears in the process of the 
polarization reversal in ferroelectrics,
it inevitably vanishes when the polarization reversal is completed. 
Quantum pumping current proposed by Thouless \cite{Thouless},
on the other hand, can support dc current, but it  
requires a nontrivial topological (winding) number 
defined in the parameter space of the Hamiltonian which is only achieved 
with a large deformation of the Hamiltonian in the parameter space. 
The Hall current in the quantum Hall effect is also such current characterized 
by a nontrivial Berry phase \cite{TKNN}. 
While the quantum Hall current is steady dc current, it is carried 
by the edge channels (not through the bulk) and is realized under the 
nontrivial topological (Chern) number which usually requires an application 
of a large external magnetic field.
Therefore, it has been considered to be
difficult to realize dc current in the presence of a band gap 
when the Hamiltonian does not possess any nontrivial topological number.

The restriction on obtaining dc current is relaxed in the non-equilibrium 
state under the light irradiation.
Of particular interest is the shift current as a mechanism of the photocurrent in 
noncentrosymmetric crystal~%
\cite{Kraut,Sipe,Young-Rappe,Young-Zheng-Rappe,Kral,Morimoto}. 
This might be relevant to the recent experiments
showing the high efficiency solar cell action~\cite{Grinberg,Nie,Shi,deQuilettes,Bhatnagar}.
Shift current is induced by
the change in the intra-cell coordinates associated with the interband transitions. 
Namely, the difference of the Berry phases 
between the conduction and valence bands induces
the steady dc current in noncentrosymmetric 
crystals even in the absence of an external dc electric field. 
However, it is assumed here that electrons and holes are independent free particles, 
i.e., the single particle approximation is employed. 

The other example of the second order optical effects is the second harmonic generation 
(SHG)~\cite{Bloembergen,Boyd,Toyozawa}. 
SHG is a common optical process which is used to detect the inversion 
symmetry breaking both in the bulk crystal and at interfaces or surfaces. 
When the incident light has the frequency $\Omega$, 
the second order nonlinear responses can have two output frequencies, i.e., 0 and $2 \Omega$. The first one corresponds to the shift current discussed above or optical rectification if it is detected optically in sufficiently low frequencies. The optical rectification is important in generating terahertz (THz) light.
The latter $2\Omega$ response corresponds to the SHG. It is experimentally shown that the excitons can contribute to 
the optical rectification~\cite{Bieler,Dawson} and the SHG~\cite{Minami,Shen,Lafrentz}.
While the optical rectification and the SHG are well-known nonlinear optical effects, their enhancement due to the exciton resonance has not been fully explored from the viewpoint of topology and geometry of the Bloch electrons.

In the present paper, we study the role of exciton formation 
on the second order optical processes and demonstrate that they are
topological in nature. We show that (i) the shift current originating 
from the Berry phase remains nonvanishing 
even when the excitons are formed due to the attractive interaction 
between an electron and a hole created by the light irradiation,
and (ii) the SHG is expressed by the similar expression to the
shift current in terms of the Berry connection and is enhanced at the exciton resonance. 
This is achieved by using the Floquet two band model developed in 
Ref.~\cite{Morimoto} by incorporating the attractive interaction. 
This formalism enables us to concisely describe nonequilibrium steady states with 
exciton formation and study various nonlinear current responses produced by exciton creation.

\begin{figure}
\begin{center}
\includegraphics[width=0.66\linewidth]{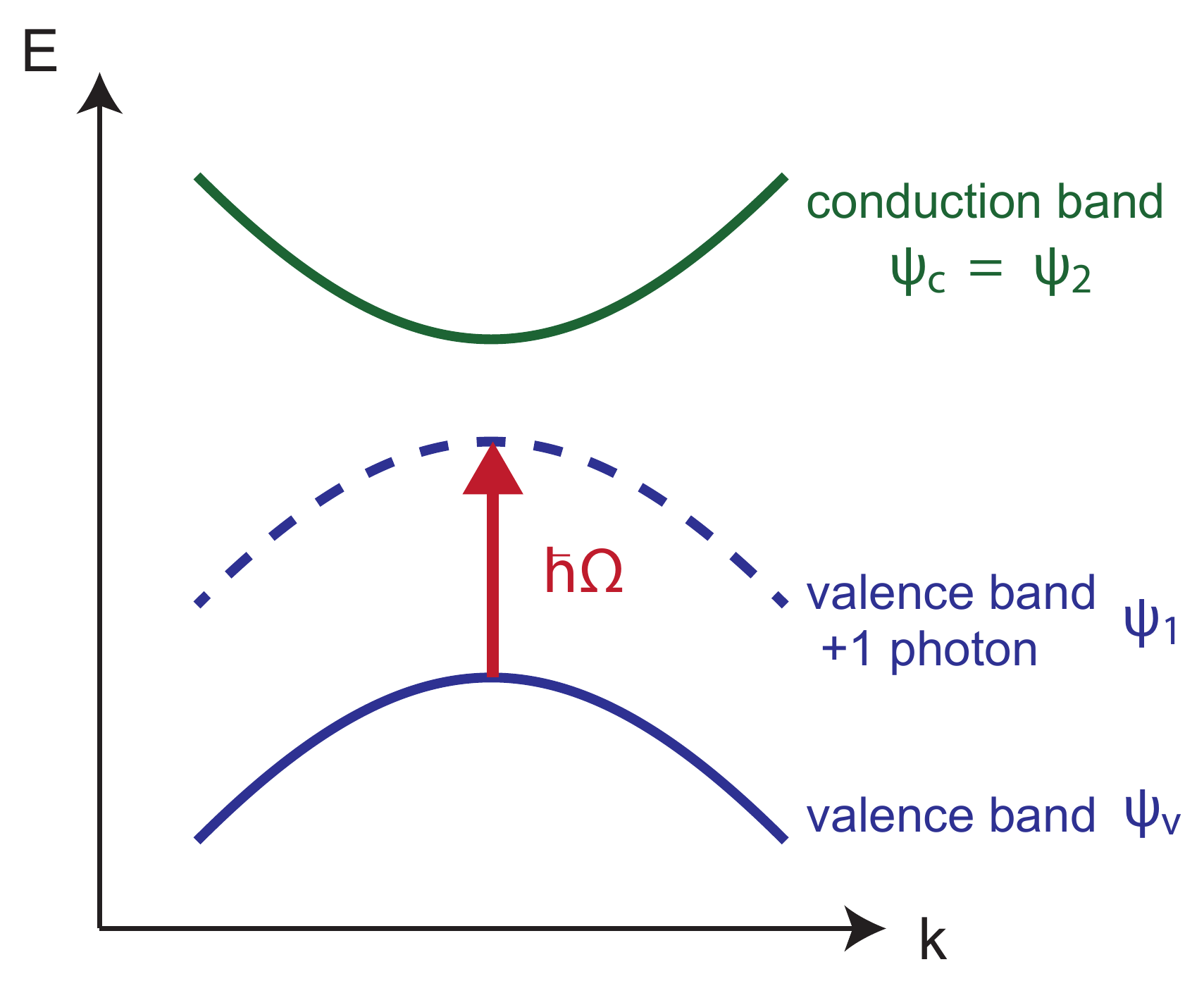}
\caption{\label{fig: floquet band}
Schematic picture of the Floquet two band model.
Bands labeled by $\psi_1$ and $\psi_2$ denote the valence band with Floquet index $-1$ and the conduction band with Floquet index $0$, respectively. Quasienergy for the Floquet band $\psi_1$ is smaller than that for $\psi_2$. The exciton formation is described by an effective mixing of these two bands in the mean field approximation of the electron-electron interaction.
}
\end{center}
\end{figure}

\section{Floquet two band model for excitons}
We study a two band model that describes the exciton formation in a system driven 
by an external electric field of light.
We consider a $d$-dimensional system in which the electric field is applied 
along the $i$th direction and the repulsive interaction is present between the valence 
electron and the conduction electron.
Then the Hamiltonian of the two band model is given (with the convention $e=\hbar=1$) by
\begin{align}
H&=\sum_{\alpha=c,v} \sum_{\bm{k}} \epsilon_i^0[\bm{k}+\mathcal A(t)\bm{e_i}] 
\psi_{\alpha,\bm{k}}^\dagger \psi^{\,}_{\alpha,\bm{k}} 
\n &\qquad
+\sum_{\bm{k}} \mathcal A(t) [v_{12}(\bm{k}) \psi_{v,\bm{k}}^\dagger 
\psi^{\,}_{c,\bm{k}} +h.c.]
\n &\qquad
- \sum_{\bm{k},\bm{k'}} V_{\bm{k}\bm{k'}}\psi_{c,\bm{k}}^\dagger 
\psi^{\,}_{v,\bm{k}} \psi_{v,\bm{k'}}^\dagger \psi^{\,}_{c,\bm{k'}},
\label{eq: H(t)}
\end{align}
where $\psi_v$ and $\psi_c$ are annihilation operators for valence and conduction bands 
with the energy dispersions 
$\epsilon_v^0(\bm{k})$ and $\epsilon_c^0(\bm{k})$, respectively.
Electrons are driven by an electric field 
$E(t)\bm{e_i}$ (with the $i$th unit vector $\bm{e_i}$) 
which is periodic in time as 
\begin{align}
E(t)=E e^{-i \Omega t}+E^* e^{i\Omega t}.
\end{align}
This electric field is
introduced to the Hamiltonian by the substitution 
$\bm{k}\to \bm{k}+\mathcal A(t)\bm{e_i}$ 
with the gauge potential is given by $\mathcal A(t)=i A e^{-i \Omega t} -i A^* e^{i\Omega t}$ with $A=E/\Omega$.
In addition to the first term in Eq.~(\ref{eq: H(t)}) that corresponds to this substitution in the energy dispersion, 
the electric field leads to an interband effect described by the second term in Eq.~(\ref{eq: H(t)}), i.e., the coupling to the current matrix element 
\begin{align}
v_{12}(\bm{k})=\bra{\psi_{v,\bm{k}}} v \ket{\psi_{c,\bm{k}}} 
\end{align}
between the valence and conduction bands where $v$ is the velocity 
operator in the $i$th direction.
The attractive interaction between electron and hole  
is described by $V_{\bm{k},\bm{k'}}$ which leads to the exciton formation.
We have picked up only the interaction terms which are
relevant to the formation of excitons with zero center-of-mass
momentum corresponding to the uniform electric field of light.
This is analogous to the BCS Hamiltonian of superconductivity, but
we do not discuss the condensate of the excitons here.

The nonequilibrium steady state under the light irradiation is concisely described 
by using the Floquet formalism combined with the Keldysh Green's function method~%
\cite{Kohler,Jauho,Johnsen,Kamenev,Oka,Hanai,Morimoto}.
The Floquet formalism offers a description of periodically driven systems in terms of Floquet bands.
Specifically, we define a Floquet Hamiltonian $H_F$ with Fourier transformation of the time dependent Hamiltonian $H_0(t)$ of the period $T$ as
\begin{align}
(H_F)_{mn}= \frac{1}{T} \int_0^T dt e^{i(m-n)\Omega t} H_0(t)
-n\Omega \delta_{mn},
\end{align} 
where $m$ and $n$ are Floquet indices and $\Omega=2\pi/T$.
While Floquet bands obtained from $H_F$ determines eigenstates in periodically driven systems, they lack the information of how they are occupied in the steady state.
The occupation of the Floquet bands in the steady state can be fixed by coupling the system to a heat bath having the Fermi energy and the temperature that we want to impose onto the system.
This is concisely described by using the Keldysh Green's function method and include the effect of the heat bath as a self energy.
The Keldysh Green's functions in the Floquet formalism are written by the Dyson equation as
\begin{align}
\begin{pmatrix}
G^R & G^K \\
0 & G^A
\end{pmatrix}^{-1}
&=\omega - H_F + \Sigma.
\label{eq: dyson eq}
\end{align}
Here we consider two contributions to the self energy as 
$\Sigma=\Sigma_\t{bath}+\Sigma_\t{ex}$,
where $\Sigma_\t{bath}$ is the self energy arising from a coupling to the bath and $\Sigma_\t{ex}$ is the self energy arising from the exciton formation due to the electron-electron interaction.
The self energy $\Sigma_\t{bath}$ for the heat bath is given by
\begin{align}
(\Sigma_\t{bath})_{mn}=i\Gamma \delta_{mn}
\begin{pmatrix}
\frac{1}{2} & -1 + f(\omega+m\Omega) \\
0 & -\frac 1 2
\end{pmatrix},
\end{align}
for Floquet indices $m$ and $n$.
This form of $\Sigma_\t{bath}$ assumes that 
each site is coupled to a heat bath which has a wide spectrum and the distribution function $f(\epsilon)$, and $\Gamma$ measures the strength of the coupling between the system and the bath~\cite{Kohler}.
The inclusion of $\Sigma_\t{bath}$ fixes the occupation of the Floquet bands properly through the Keldysh component of the Dyson equation.
The self energy $\Sigma_\t{ex}$ describes the exciton formation in the driven system which we incorporate in the mean-field approximation for the interaction 
term by keeping the Fock term in the Keldysh Green's function.

Now we apply this formalism to the two band model in Eq.~(\ref{eq: H(t)}).
When the electric field is weak, we can focus on two Floquet bands, i.e., 
the valence band dressed with one photon and the conduction band dressed with zero photon, 
which are denoted by annihilation operators $\psi_1$ and $\psi_2$, respectively, 
as schematically illustrated in Fig.~\ref{fig: floquet band}.
Here, subscripts 1 and 2 are shorthands for the valence band with Floquet index $-1$ and the conduction band with Floquet index $0$, respectively.
In this case, the Floquet Hamiltonian is given by \cite{Morimoto}
\begin{align}
\mathcal{H}_F&=
\begin{pmatrix}
\psi_{1,\bm{k}}^\dagger & \psi_{2,\bm{k}}^\dagger
\end{pmatrix}
H_F
\begin{pmatrix}
\psi_{1,\bm{k}} \\ \psi_{2,\bm{k}}
\end{pmatrix}
\\
H_F&=
\begin{pmatrix}
\epsilon_1(\bm{k}) & -iA^* [v_{12}(\bm{k}) + v'^*(\bm{k})] \\
iA [v_{21}(\bm{k})+v'(\bm{k})] & \epsilon_2(\bm{k})
\end{pmatrix}
\n
&\equiv
\epsilon+\Vec d \cdot \Vec \sigma,
\label{eq: HF1}
\end{align}
where
$\epsilon_1=\epsilon_v^0+\hbar \Omega$,
$\epsilon_2=\epsilon_c^0$,
and
$v_{21}=v_{12}^*$.
The detuning
$d_z(\bm{k})=-\frac{1}{2}[\epsilon^0_c(\bm{k})-\epsilon^0_v(\bm{k})-\hbar\Omega]$ 
is negative for any value of $\bm{k}$, 
because we are interested in the excitonic bound state
where the photon energy is smaller than the band gap.
The nonzero expectation value of excitons effectively modifies the 
dipole matrix element by
\begin{align}
i A v'(\bm{k})&=-\int d\bm{k'} V_{\bm{k}\bm{k'}}\Delta(\bm{k'}),
\\
\Delta(\bm{k})&=\braket{\psi_{1,\bm{k}}^\dagger \psi^{\,}_{2,\bm{k}}}.
\end{align}
This mean field treatment of excitons in $H_F$ is equivalent to including the retarded component of the exciton self energy $\Sigma_\t{ex}^R$ into the mean field Floquet Hamiltonian $H_F$ in the Dyson equation [Eq.~(\ref{eq: dyson eq})].

The exciton formation is captured by the self-consistency equation for 
$\Delta(\bm{k})$ which we solve by employing the Keldysh Green's function in the following.
First, the lesser Green's function for the Floquet two band model is given by \cite{Morimoto}
\begin{align}
G^<&= G^R \Sigma^< G^A
\n
&=
\frac{(\omega-\epsilon-i\Gamma+\Vec d \cdot \Vec \sigma)
\Sigma^<
(\omega-\epsilon+i\Gamma+\Vec d \cdot \Vec \sigma)}
{[(\omega-\epsilon-i\Gamma)^2-d^2][(\omega-\epsilon+i\Gamma)^2-d^2]},
\end{align}
with 
\begin{align}
\Sigma^<=\frac{\Sigma^R+\Sigma^K-\Sigma^A}{2}=i\Gamma \frac{1+\sigma_z}{2}.
\label{eq: Sigma<}
\end{align}
Here the lesser self-energy $\Sigma^<$
 describes the occupation of Floquet bands and is determined by the heat bath as $\Sigma^< \cong \Sigma_\t{bath}^<$. 
The above form of $\Sigma^<$ assumes that the Fermi energy is located within the energy gap of the original band structure \cite{Kohler}.
Specifically, 
the final equation in Eq.~(\ref{eq: Sigma<}) follows from $f(\epsilon_v^0)=1$ and $f(\epsilon_c^0)=0$ since 
$(\Sigma_\t{bath}^<)_{mn}=i\Gamma \delta_{mn} f(\omega+m\Omega)$.
Next, in the case of two band model, general expectation values $\braket{\psi^\dagger (\bm{b}\cdot \bm{\sigma})^T \psi}$ for any $\bm b=(b_x,b_y,b_z)$
can be evaluated by using the above lesser Green's function as \cite{Morimoto}
\begin{align}
\braket{\psi^\dagger (\bm{b}\cdot \bm{\sigma}) \psi} 
&=
-i\t{Tr}[G^< (\bm{b}\cdot \bm{\sigma})] \n
&=
\int d\bm{k} 
\frac{1}{d^2+\frac{\Gamma^2}{4}}
\Bigg[
\frac{\Gamma}{2}(-d_x b_y+d_y b_x)
\n &\qquad
+(d_x b_x+d_y b_y)d_z
+(d_z^2+\frac{\Gamma^2}{4}) b_z
\Bigg]
\label{eq:exp}
,
\end{align}
where $\t{Tr}$ denotes the trace of a matrix and integration over $\bm{k}$ and $\omega$.
The self-consistency condition for $\Delta(\bm{k})$ 
is written by using the above equation with 
$\bm{b}\cdot \bm{\sigma}=(\sigma_x+i\sigma_y)/2$
as 
\begin{align}
\Delta(\bm{k})&=  \braket{\psi_{1,\bm{k}}^\dagger \psi^{\,}_{2,\bm{k}}}
=-i \int d \omega  G^<_{21} 
\n 
&=
\frac{
iA(v_{21}+v')(d_z-i\frac{\Gamma}{2})
}{2(d^2+\frac{\Gamma^2}{4})},
\end{align}
which is essentially equivalent to the Dyson equation for the retarded 
component of the self energy $\Sigma^R$ \cite{Hanai}.
This leads to the integral equation,
\begin{align}
\Delta(\bm{k}) &=
i A v_{21} \frac{
d_z-i\frac{\Gamma}{2}
}{2(d^2+\frac{\Gamma^2}{4})}
-
\frac{
d_z-i\frac{\Gamma}{2}
}{2(d^2+\frac{\Gamma^2}{4})}
\int d\bm{k'} V_{\bm{k}\bm{k'}} \Delta(\bm{k'})
.
\label{eq: self consistent eq}
\end{align}
If we assume that the attractive interaction has the separable form 
\begin{align}
V_{\bm{k}\bm{k'}}=w^*(\bm{k})w(\bm{k'}),
\end{align}
we can solve the integral equation 
as
\begin{align}
v'(\bm{k})&=-w^*(\bm{k})B, &
B&=\frac{1}{i A}\int d\bm{k'} w(\bm{k'})\Delta(\bm{k'}),
\end{align}
where the integral equation for $\Delta(\bm k)$ reduces to the linear equation for $B$ 
given by
\begin{align}
B&=
\int d\bm{k} w(\bm{k}) v_{21} \frac{
d_z-i\frac{\Gamma}{2}
}{2(d^2+\frac{\Gamma^2}{4})}
-
\int d\bm{k} |w(\bm{k})|^2
\frac{
d_z-i\frac{\Gamma}{2}
}{2(d^2+\frac{\Gamma^2}{4})}
B
.
\end{align}
When $A|v_{21}+v'|$ is much smaller than $|d_z|$ and $\Gamma$
(i.e., the external electric field is not too strong),
$v'$ is written as
\begin{align}
v'(\bm{k})&=-w^*(\bm{k})\frac{C_1}{1+C_2},
\end{align}
with 
\begin{align}
C_1&=\int d\bm{k} \frac{
 w(\bm{k}) v_{21}
}{2(d_z+i\frac{\Gamma}{2})}, 
&
C_2&=\int d\bm{k} 
\frac{
|w(\bm{k})|^2
}{2(d_z+i\frac{\Gamma}{2})}
.
\label{eq: C1 and C2}
\end{align}
Intuitively, $1/(1+C_2)$ corresponds to the propagator of the exciton,
and the resonance to the exciton state takes place when $\t{Re}(1+C_2)=0$ is satisfied 
by the incident light frequency $\Omega$.
In particular, when the detuning $d_z(\bm{k})$ is constant as a function of $\bm{k}$ 
(as in flat bands) and the interaction is of a contact type 
[i.e., $w(\bm{k})$ is a constant satisfying $\int d\bm{k} |w(\bm{k})|^2=V$],
the exciton resonance takes place at the frequency
$\hbar\Omega=\epsilon^0_c-\epsilon^0_v-V$.
In the following, we show that the shift current is nonvanishing 
in the presence of the exciton formation where no free electrons and holes are created.

Now we study the current 
$J\equiv \braket{\psi^\dagger \partial_{k_j} H_F \psi}$ in the $j$th direction 
in the presence of exciton formation.
The current expectation value is obtained by setting 
$b_x-ib_y=-i A^* (\partial_{k_j} v)_{12}$ and $b_z=(\partial_{k_j} \epsilon_1
-\partial_{k_j} \epsilon_2)/2$ in 
Eq.~(\ref{eq:exp}),
which gives
\begin{align}
J&=\int d\bm{k}
(j_1+j_2),
\label{eq: J}
\end{align}
with
\begin{align}
j_1&=
\int d\bm k \frac{
\t{Re}[(d_z-i\frac{\Gamma}{2})(d_x+i d_y)(b_x-ib_y)]
}{d^2+\frac{\Gamma^2}{4}}
\n
&=
|A|^2\frac{
\t{Re}\{(d_z-i\frac{\Gamma}{2})[(\partial_{k_j} v)_{12} (v_{21}+v')] \}
}{d^2+\frac{\Gamma^2}{4}} ,
\label{eq: j1}
\\
j_2&=
\frac{
(d_z^2+\frac{\Gamma^2}{4})(\partial_{k_j} \epsilon_1-\partial_{k_j} \epsilon_2)
}{2(d^2+\frac{\Gamma^2}{4})}.
\label{eq: j2}
\end{align}
When $|A(v_{21}+v')|$ and $\Gamma$ are much smaller than $|d_z|$,
we can replace $d^2$ with $d_z^2$ in the denominators. 
Then $j_2$ vanishes after the integration over $\bm{k}$ because 
$\int d\bm{k} \partial_{k_j} \epsilon_\alpha=0$;
we focus on the contribution from $j_1$ hereafter.
In the two band model, the derivative of the velocity operator in Eq.~(\ref{eq: j1}) 
is written as \cite{Morimoto}
\begin{align}
\left(\frac{\partial v}{\partial k_j} \right)_{12}
&=\frac{\partial  v_{12}}{\partial k_j}
-\bra{\partial_{k_j} u_1} v \ket{u_2}
-\bra{u_1}v \ket{\partial_{k_j} u_2} \n
&=v_{12} (R_1+iR_2),
\end{align}
with
\begin{align}
R_1&=\partial_{k_j} \log |v_{12}| +\frac{v_{11}- v_{22}}{\epsilon_1-\epsilon_2}, \\
R_2&=\partial_{k_j} \t{Im}[\log v_{12}] + a_{1}-  a_{2}, 
\end{align}
Here, $u_\alpha$ is the periodic part of the Bloch wave function
and  $a_{\alpha}=-i \bra{u_\alpha} \partial_{k_j} u_\alpha  \rangle$ 
is the Berry connection of the band $\alpha$.
We note that $R_1$ and $R_2$ have dimensions of the length. In particular, 
$R_2$ is known as the shift vector
and describes the shift of the wavepackets in the valence and conduction bands.
Intuitively, the shift vector $R_2$ is $\bm k$-resolved version of electric polarization and originates from the difference of intra-cell coordinates for the valence 
and conduction bands which is expressed by the Berry connections.
Indeed, the $\bm k$-integral of $R_2$ is the difference of electric polarizations of the valence and conduction bands as can be seen from 
\begin{align}
\int d\bm k R_2=\int d \bm k a_1 -\int d \bm k a_2,  
\end{align}
where the contribution of $\partial_{k_j} \t{Im}[\log v_{12}]$ vanishes 
because it is a total derivative with respect to $k_j$.
In the presence of the time reversal symmetry (TRS),
$R_1$ and $R_2$ are odd and even in $\bm{k}$, respectively,
and $|v_{12}|^2$ is even in $\bm{k}$.
Thus, the photocurrent from the excitons in Eq.~(\ref{eq: J}) 
reduces in the presence of TRS to
\begin{subequations}
\label{eq: J TRS}
\begin{align}
J&= J_\t{con} + J_\t{ex}, 
\\
J_\t{con} &= |A|^2 \int d\bm{k} 
\frac{\frac{\Gamma}{2}}{d_z^2+\frac{\Gamma^2}{4} } |v_{12}|^2 R_2,
\\
J_\t{ex} &=
|A|^2 \int d\bm{k} \frac{1}{d_z} 
[
\t{Re}(v_{12} v') 
R_1
-
\t{Im}(v_{12} v') R_2
]
.
\end{align}
\end{subequations}
The first term $J_\t{con}$ describes the conventional shift current that 
involves creation of a pair of free electron and hole which is present for 
$\hbar \Omega > E_g$ with the band gap $E_g$.
The second term $J_\t{ex}$ describes the shift current carried by excitons and is 
nonvanishing even when $\hbar \Omega <E_g$. We note that we dropped 
$\Gamma$ in the second term because we can assume 
$\Gamma \ll |d_z|$ in describing excitons.

We study properties of the exciton photocurrent $J_\t{ex}$ in the following.
First, the photocurrent $J_\t{ex}$ is generated through the real transitions to create excitons,
because only the imaginary part of the exciton propagator $1/(1+C_2)$ 
contributes to the photocurrent $J_\t{ex}$ in Eq.~(\ref{eq: J TRS}).
This is reasonable from the viewpoint of the energy conservation.
It is easy to explicitly show this fact in the presence of the TRS
by assuming that the time reversal operation is represented by a 
complex conjugation ($T=\mathcal{K}$).
In this case, the velocity operator obeys the equation 
$v_{ij}(-\bm{k})=-(v_{ij}(\bm{k}))^*$ and 
the separable interaction term $w(\bm{k})$ can be chosen to satisfy 
$w(-\bm{k})=w^*(\bm{k})$ without loss of generality.
Since the real part of $w(\bm{k})v_{21}(\bm{k})$ is odd in $\bm{k}$,
$C_1$ is pure imaginary at the exciton resonance 
where we can drop $\Gamma$ in the denominator.
By noticing that $w^*(-\bm{k})v_{12}(-\bm{k})=-[w^*(\bm{k})v_{12}(\bm{k})]^*$,
we can write the photocurrent as
\begin{align}
J_\t{ex}
&=
|A|^2 \int d\bm{k} \frac{\t{Im}(C_1)}{d_z}
\t{Im}\left[\frac{1}{1+C_2}\right]
\n&\qquad\times 
\left\{
\t{Re}[w^*(\bm{k})v_{12}(\bm{k})] 
R_1
-
\t{Im}[w^*(\bm{k})v_{12}(\bm{k})] R_2
\right\},
\label{eq: Jex with TRS}
\end{align}
which is proportional to $\t{Im}\left[1/(1+C_2)\right]$ and manifests that the 
photocurrent is generated by real transition to the exciton state.
Second, the expression for $J_\t{ex}$ can be further simplified for shallow excitons.
Shallow excitons are those with small binding energy that are formed by Bloch states near the band gap.
Specifically, in this case of shallow excitons, the $\bm{k}$-integrals are contributed only from the small region $(\delta k)^d$ 
around $\bm{k}=\bm{0}$ where the band gap is the smallest;
we replace $\int d\bm k$ with $\int d\bm k (\delta k)^d \delta(\bm k)$ in 
Eq.~(\ref{eq: Jex with TRS}).
By doing so, the mean field solution of the exciton in Eq.~(\ref{eq: C1 and C2}) leads to
\begin{align}
\t{Im}[C_1]&=(\delta k)^d \frac{w(0)}{2d_z(0)} \t{Im}[v_{21}(0)], \\
\t{Im}\left[\frac{1}{1+C_2}\right]&=-2d_z(0) \frac{\Gamma}{[2d_z(0)+V']^2 + \Gamma^2}, 
\end{align}
with  $V'=|w(0)|^2 (\delta k)^d$.
Thus the photocurrent $J_\t{ex}$ for shallow excitons is given by
\begin{align}
J_\t{ex} &\cong 
|A|^2\frac{V' (\delta k)^d}{|d_z(0)|} \frac{\Gamma}{[2d_z(0)+V']^2+\Gamma^2}
|v_{12}(0)|^2 R_2(0)
,
\label{eq: Jex shallow exciton}
\end{align}
where we used the relations 
$\t{Im}[v_{12}(0)]\t{Im}[v_{21}(0)]=-|v_{12}(0)|^2$ and $d_z<0$.
In this expression, the factor $\Gamma/\{[2d_z(0)+V']^2 + \Gamma^2 \}$ 
describes the real transition into the exciton state at the resonance frequency
 $\hbar\Omega=\epsilon^0_c-\epsilon^0_v-V'$.
In addition, the shift current for shallow excitons is proportional to the weight of the exciton 
$V' (\delta k)^d/[2|d_z(0)|]$.
This clearly shows that the nonzero shift current flows by 
creating excitons below the band gap.
Furthermore, we notice that $J_\t{ex}$ is proportional to the contribution to the conventional shift current $J_\t{con}$ 
at $\bm k = \bm 0$ that has a factor $|v_{12}(0)|^2 R_2(0)$ as seen in Eq.(\ref{eq: J TRS}b).
This indicates that the conventional shift current in the noninteracting system is partly
transfered to the exciton resonance below the band gap due to the exciton formation 
with the attractive interaction. 
In fact, by recovering the energy broadening $\Gamma$ in Eq.~(\ref{eq: J TRS}c), one obtains the term $|A|^2 \frac{\Gamma/2}{d_z^2+\Gamma^2/4}\t{Im}[(\partial_{k_j}v)_{12} v']$ which gives negative shift current contribution for the electron-hole continuum as follows.
For simplicity, we focus on the case where the resonance condition is satisfied at the band gap at $k = 0$ (i.e., $d_z(0)=0$ and $d_z (k)\neq 0$ for $k\neq 0$). In this case, Eq.~(\ref{eq: C1 and C2}) gives 
$C_1=-i \pi w(0) v_{21}(0) D(0)$ and $C_2=- i \pi |w(0)|^2 D(0)$ 
where $D(0)$ is the joint density of states at $k=0$, and the above term is expressed as 
$|A|^2 \frac{\Gamma/2}{d_z^2+\Gamma^2/4} |v_{12}(0)|^2 R_2(0) [- \frac{\pi |w(0)|^2 D(0)}{1+ (\pi |w(0)|^2 D(0))^2}]$.
This should be compared with $J_\t{con}$ and clearly describes the partial suppression of the conventional shift current above the band gap due to the exciton formation.

An optical process that is closely related to the shift current is optical rectification. The optical rectification is the second order nonlinear optical effect that optically measures emission of low frequency light, typically in the THz regime. Namely, the optical rectification is a low frequency optical analog of the shift current and is important for application for THz generation.
Since the shift current is enhanced at the exciton resonance below the band gap, the optical rectification is also enhanced at the exciton resonance. Thus strong THz generation is expected by shining the light to noncentrosymmetric crystals at the exciton resonance. Indeed, there are experimental reports on enhanced THz emissions for GaAs when the laser frequency is resonant to the excitons \cite{Bieler,Dawson}.

\section{Second Harmonic Generation}
Exciton formation also enhances the SHG for the photon energy below the band gap in a similar manner to the case of shift current.
The SHG is the current response of the frequency $2\Omega$
when the incident light has the frequency $\Omega$.
In our formalism, the SHG can be studied by using the formula for time-dependent current,
\begin{align}
J(t)= -i \sum_m \t{Tr}[v(t) G^<_{mn}] e^{-i(m-n)\Omega t},
\label{eq: J(t)}
\end{align} 
where subscripts $m$ and $n$ denote the Floquet indices, and 
$\t{Tr}$ denotes a trace over the band indices and $\omega$ and $\bm k$ integration. 
Here the time-dependent current operator is given by
\begin{align}
v(t)&=v + (i A e^{-i\Omega t} \partial_k v + \t{h.c.}) +O(A^2).
\end{align}
The frequency $2\Omega$ component of the current $J_{2\Omega}$ is decomposed into two contributions as 
\begin{align}
J_{2\Omega}=J_\t{1ph}+J_\t{2ph},
\end{align}
where the first term and the second term represent one-photon contribution and the two-photon contributions, respectively.
In the standard perturbation theory \cite{Sipe},
the one-photon contribution corresponds to the bubble diagram where the diamagnetic current is induced by the external electric field coupling to the usual current operator,
while the two-photon contribution corresponds to the diagram where the usual current response is induced by the external electric field coupling to the diamagnetic current.
In the following, we compute the one-photon contribution and the two-photon contribution separately.

The one-photon contribution $J_\t{1ph}$ is obtained by setting $m=n+1$ and 
$v(t) \to iA e^{-i\Omega t} \partial_k v$ in Eq.~(\ref{eq: J(t)}).
Since this is the same Floquet two band model as in Eq.~(\ref{eq: HF1}),
we can compute $J_\t{1ph}$ in a similar way to the shift current. In particular, the same self-consistent equation Eq.~(\ref{eq: self consistent eq}) holds for the exciton formation.
By using the formula for the lesser Green's function \cite{Morimoto},
\begin{align}
(G^<)_{21}&=\frac{(d_x+id_y)(\frac{\Gamma}{2}+id_z)}{2(d^2+\frac{\Gamma^2}{4})},
\label{eq: G<}
\end{align}
the one-photon contribution is written as
\begin{align}
J_\t{1ph}
&=
-A^2 \int d \bm k \frac{1}{2d_z} \t{Re}[(\partial_k v)_{12} v' ] \n
&=
-A^2 \int d\bm{k} \frac{\t{Im}(C_1)}{2d_z}
\t{Im}\left[\frac{1}{1+ C_2}\right]
\n&\qquad\times 
\left\{
\t{Re}[w^*(\bm{k})v_{12}(\bm{k})] 
R_1
-
\t{Im}[w^*(\bm{k})v_{12}(\bm{k})] R_2
\right\},
\label{eq: J1ph}
\end{align}
where we only kept terms relevant to the exciton resonance.
Thus the one-photon contribution is the same as the shift current $J_\t{ex}$ except that it has the factor $-A^2/2$ instead of $|A|^2$ in Eq.~(\ref{eq: Jex with TRS}).

The two-photon contribution $J_\t{2ph}$ is obtained by setting $m=n+2$ and $v(t) \to v$ in Eq.~(\ref{eq: J(t)}).
Therefore, the two photon contribution arises from another two band model in which valence and conduction bands are separated by two Floquet indices. This is given by
\begin{align}
\tilde{\mathcal{H}}_F &=
\begin{pmatrix}
\tilde \psi_1^\dagger & \tilde \psi_2^\dagger 
\end{pmatrix}
\tilde H_F
\begin{pmatrix}
\tilde \psi_1 \\
\tilde \psi_2
\end{pmatrix},
\\
\tilde H_F 
&=
\begin{pmatrix}
\epsilon_v^0 +2\hbar\Omega & -\frac 1 2 A^2 [(\partial_k v)_{12} + (\partial_k \tilde v')^*] \\
-\frac 1 2 A^2 [(\partial_k v)_{21}+ \partial_k \tilde v'] & \epsilon_c^0
\end{pmatrix},
\label{eq: tilde HF}
\end{align}
where $\tilde \psi_1, \tilde \psi_2$ are annihilation operators for the valence band with Floquet index $-2$ and the conduction band with Floquet index $0$, respectively, and $(\partial_k v)_{12}= \bra{\psi_{v, k}} \partial_k v \ket{\psi_{c, k}}$.
The off-diagonal term originate from the time-dependent Hamiltonian expanded up to the order of $A^2$.
Specifically, when we keep terms up to the order of $A^2$,
the time-dependent Hamiltonian reads
\begin{align}
H(t)=H_0 + \mathcal A(t) v + \frac{1}{2} \mathcal A(t)^2 \partial_k v,
\end{align}
and the Fourier components of $e^{\pm i 2\Omega t}$ produces the off-diagonal terms in Eq.~(\ref{eq: tilde HF}).
In this case, the self consistent equation is given by
\begin{align}
-\frac{1}{2}A^2 \partial_k \tilde v' 
&=-\int d \bm{k'} V_{\bm k \bm{k'}} \tilde \Delta(\bm{k'}), \\
\tilde \Delta(\bm{k})&= 
\braket{\tilde \psi_{1,\bm k}^\dagger \tilde \psi_{2,\bm k}} \n
&=
\frac{
-(1/2)A^2 [(\partial_k v)_{21}+ \partial_k \tilde v'] (d_z-i\frac{\Gamma}{2})
}{2(d^2+\frac{\Gamma^2}{4})}.
\end{align}
These equations describe excitons formed by two photon absorption and are different from Eq.~(\ref{eq: self consistent eq}) for the excitons formed by one photon absorption.
The self consistent equation is solved in a similar manner by assuming the separable form for the interaction 
$V_{\bm k \bm{k'}}=w^*(\bm k) w(\bm{k'})$ as
\begin{align}
\partial_k \tilde v'(\bm{k})&=-w^*(\bm{k})\frac{\tilde C_1}{1+\tilde C_2},
\end{align}
with 
\begin{align}
\tilde C_1&=\int d\bm{k} \frac{
 w(\bm{k}) (\partial_k v)_{21}
}{2(d_z+i\frac{\Gamma}{2})}
=\int d\bm{k} \frac{
 w(\bm{k}) v_{21}
}{2(d_z+i\frac{\Gamma}{2})} (R_1 -i R_2),
\\
\tilde C_2&=\int d\bm{k} 
\frac{
|w(\bm{k})|^2
}{2(d_z+i\frac{\Gamma}{2})}
.
\label{eq: tilde C1 and tilde C2}
\end{align}
Here we used the identity 
$(\partial_k v)_{21}=[(\partial_k v)_{12}]^*=[v_{12}(R_1+iR_2)]^*=v_{21}(R_1-iR_2)$.
Then we obtain the two photon contribution as
\begin{align}
J_\t{2ph}
&=
-\frac{A^2}{2} \int d \bm k \frac{1}{2d_z} \t{Re}[v_{12} (\partial_k \tilde v')_{21}] \n
&=
-A^2 \int d\bm{k} \frac{\t{Re}(\tilde C_1)}{4d_z}
\t{Im}\left[\frac{1}{1+ \tilde C_2}\right]
\t{Im}[w^*(\bm{k})v_{12}(\bm{k})]
,
\label{eq: J2ph}
\end{align}
where we only kept the term relevant to the exciton resonance.
When we equate the first line and the second line, 
we used constraints from the TRS.
Specifically, the TRS ($T=\mathcal{K}$) requires $w^*(\bm k)v_{12}(\bm k)=-[w^*(\bm -k)v_{12}(\bm -k)]^*$, $w(\bm k) \partial_k v(\bm k)=[w(\bm -k) \partial_k v(\bm -k)]^*$. 
Thus $\tilde C_1$ is real when we neglect $i\frac \Gamma 2$ in the denominator, and $\t{Re}[w^*(\bm{k})v_{12}(\bm{k})]$ is odd in $\bm k$ and vanishes after $\bm k$-integration.
Since this expression shows $J_\t{2ph} \propto \t{Im}[1/(1+ \tilde C_2)]$,
we again find that the two photon contribution to SHG is generated by the real transition to the exciton state.

Next we study SHG in the case of shallow excitons.
We assume that the integral is contributed near $\bm k=\bm 0$ due to the factor $1/d_z$, and replace $\int d \bm k$ with $\int d \bm k (\delta k)^d \delta(\bm k)$.
The one-photon contribution $J_\t{1ph}$ reduces to $-1/2$ times Eq.~(\ref{eq: Jex shallow exciton}).
In the case of the two-photon contribution, 
the self consistent solution reduces in the shallow exciton limit to
\begin{align}
\t{Re}[\tilde C_1]&=
(\delta k)^d \frac{w(0)}{2d_z(0)} \t{Im}[v_{21}(0)]R_2(0),
\\
\t{Im}\left[ \frac{1}{1+\tilde C_2} \right]
&=-2\pi d_z(0) \delta[2d_z(0)+V'],
\end{align}
Here we used $\t{Re}[v_{21}(0)]=R_1(0)=0$ under the TRS in the first line and took $\Gamma \to 0$ limit in the second line.
By using these equations in Eq.~(\ref{eq: J2ph}),
the two photon contribution is written as
\begin{align}
J_\t{2ph} &\cong
-\pi A^2 \frac{V'(\delta k)^d}{4d_z(0)}|v_{12}(0)|^2 R_2(0) \delta[2d_z(0)+V'].
\end{align}
Combining these two contributions, we obtain the SHG from shallow excitons as
\begin{align}
J_{2\Omega}
&\cong
\pi A^2 V'(\delta k)^d |v_{12}(0)|^2 R_2(0)
\n
& \times
\left[ 
\frac{\delta(\epsilon_v^0- \epsilon_c^0+ V'+ \hbar \Omega)}{2 (\epsilon_v^0- \epsilon_c^0+\hbar \Omega)}
-
\frac{\delta(\epsilon_v^0- \epsilon_c^0+ V'+2 \hbar \Omega)}{4 (\epsilon_v^0- \epsilon_c^0+\hbar \Omega)}
\right]
.
\label{eq: SHG shallow exciton}
\end{align}
This clearly shows that the SHG is enhanced with exciton formation when the photon energy $\hbar \Omega$ is the same as or the half of the exciton creation energy ($\epsilon_c^0- \epsilon_v^0 - V'$).

Finally we comment on the relationship between the SHG and the shift current.
Let us define nonlinear conductivities for SHG and shift current as
\begin{align}
J_{2\Omega}&=\sigma^{(2)}(\Omega) E(\Omega)^2, \\
J&=\sigma^{(0)}(\Omega) |E(\Omega)|^2. 
\end{align}
In the case of noninteracting systems, the real part of the nonlinear conductivity for SHG is related to that for shift current as \cite{Morimoto}
\begin{align}
\t{Re}[\sigma^{(2)}(\Omega)]=-\frac{1}{2} \sigma^{(0)}(\Omega)+ \frac{1}{4}\sigma^{(0)}(2\Omega).
\end{align}
This is obtained by replacing $v'$ with $v$ in expressions for SHG [Eq.~(\ref{eq: J1ph}) and Eq.~(\ref{eq: J2ph})] and comparing it with $J_\t{con}$.
While the above relation still holds for the one-photon contribution with exciton formation, i.e., $J_\t{1ph}=-\frac{1}{2} J_\t{ex}$,
the two photon contribution does not satisfy this relation because the mean-field solution for the two-photon contribution involves the $\bm k$-integral of $\partial_k v$ in $\tilde C_1$ in contrast to $v$ in $C_1$.
However, this relationship recovers in the case of shallow excitons as is noticed by comparing Eq.~(\ref{eq: Jex shallow exciton}) and Eq.~(\ref{eq: SHG shallow exciton}). 
Thus the SHG and the shift
current are closely related with each other even in the
presence of exciton formation, and both are governed
by the shift vector $R_k(\bm k)$ which is essentially
a topological quantity described by
Berry connections.
Since the $\bm k$-integral of $R_k(\bm k)$ over the Brillouin zone coincides with the difference of polarizations of valence and conduction bands, both SHG and shift current are considered to be topological phenomena akin to electric polarization phenomena in ferroelectric materials.

\section{Discussions}
We have shown that the excitons can produce shift current under the steady light irradiation.
The absence of the inversion symmetry, i.e., the noncentrosymmetric crystal structure, is
essential for this effect, since otherwise the two contributions from $\bm{k}$ and
$-\bm{k}$ cancel each other as discussed in Ref.~\cite{Morimoto}. 
In addition, the experimental test of the prediction in the present paper requires
(i) well-defined exciton absorption peak separated from the electron-hole 
continuum, (ii) low enough temperature to suppress the thermal dissociation of
excitons into electrons and holes, (iii) well-separated electrodes from the light
irradiation spot to eliminate the contribution from the exciton dissociation at 
electrodes.  
It is also mentioned here that the shift current of excitons 
can be generalized to that of spin waves in noncentrosymmetric magnets, e.g.,
the electromagnons in chiral magnets.

Shift current of excitons can be also detected in optical measurements. 
When the incident light has two frequencies $\Omega_1$ and 
$\Omega_2$, the second order nonlinear effect allows that two harmonics $\Omega_1 - \Omega_2$ and
$\Omega_1 + \Omega_2$ are generated. 
In particular, the former one is used to 
generate the THz light~\cite{Bieler,Dawson},
and corresponds to the shift current when $\Omega_1 = \Omega_2 = \Omega$.
Therefore, by tuning $\Omega_1$, $\Omega_2$, one can see whether the
current remains finite in the limit of $\Omega_1 - \Omega_2 \to 0$. This offers 
an experimental test of the dc shift current without the complications related to the contact to the leads.
Furthermore, this indicates that the THz generation and the SHG are enhanced when the incident light is resonant to the exciton state below the band gap.
One example of noncentrosymmetric materials to study such nonlinear optical effects of excitons would be transition metal dichalcogenide monolayers such as MoS$_2$, because MoS$_2$ monolayers are noncentrosymmetric and known to show strong exciton binding~\cite{Splendiani,Mak,Eda}.

A comment is in order for the mechanism of relaxations.
In our model, the relaxation originates from the fact that each site is coupled to 
a heat bath with a fixed distribution function.
This is introduced by the self energy $\Sigma$ and realizes the nonequilibrium 
steady state with finite shift current. 
Here, it is assumed that the exchange of electrons between the system and the heat bath 
does not lead to a change in polarization.
In contrast, the recombination of electron-hole pairs (which is also a source of relaxation) 
results in a decrease in polarization 
and reduces the shift current.
Therefore, the shift current from the exciton formation requires a relaxation process 
which involves no change in polarization
and whose efficiency is larger than the recombination process.
For example, this requirement is satisfied by an isotropic heat bath such as a partially 
filled band that can exchange charge degrees of freedom with the two band system 
involved in the exciton formation.

Finally, the physical picture of the exciton shift current is sketched.
The exciton formation results in the polarization due to the shift between a 
hole in the valence band and an electron in the conduction band which is 
quantified by the Berry phase.
When excitons are constantly created in the nonequilibrium situation,
the continuous increase of the polarization in time produces 
the steady dc current. 
This mechanism is analogous to the quantum Ratchet motion in the presence of the 
asymmetry, and in sharp contrast to the charge pumping in the ground state. 
In the latter case, large amplitude deformation of the Hamiltonian 
is required to achieve a nontrivial winding number; 
in the quantum Rachet motion, only a small amplitude oscillation of a parameter 
in the Hamiltonian is sufficient
to support the constant dc current and energy supply.
Therefore, the nonequilibrium states will offer a new avenue for the physics of Berry phase.

\begin{acknowledgements}
We thank enlightening discussions with Y. Tokura, Y. Nakamura, 
M. Kawasaki, J. E. Moore, J. Orenstein, and B. M. Fregoso.
This work was supported by the EPiQS initiative of the
Gordon and Betty Moore Foundation (TM), and by JSPS
Grant-in-Aid for Scientific Research (Grants No. 24224009
and No. 26103006) from MEXT, Japan, and ImPACT Program
of Council for Science, Technology and Innovation (Cabinet
office, Government of Japan) (NN).

\end{acknowledgements}

\bibliography{exciton.bib}

\end{document}